\documentclass[twocolumn]{aastex6}

\usepackage{graphicx}

\newcommand{\lsim}{\raisebox{-.5ex}{$\,\stackrel{\textstyle <}{\sim}\,$}}
\newcommand{\gsim}{\raisebox{-.5ex}{$\,\stackrel{\textstyle >}{\sim}\,$}}

\begin{document}

\title{
Discovery of a new stellar sub-population residing in the (inner) stellar halo of the Milky Way
}

\author{
Jos\'e G. Fern\'andez-Trincado\altaffilmark{1,2},
Timothy C. Beers\altaffilmark{3}, 
Vinicius M. Placco\altaffilmark{3},
Edmundo Moreno\altaffilmark{4},
Alan Alves-Brito\altaffilmark{5},
Dante Minniti\altaffilmark{6,7,8},
Baitian Tang\altaffilmark{9},
Angeles P\'erez-Villegas\altaffilmark{10},
C\'eline Reyl\'e\altaffilmark{2},
Annie C. Robin\altaffilmark{2},
and
Sandro Villanova\altaffilmark{11}
}

\altaffiltext{1}{Instituto de Astronom\'ia y Ciencias Planetarias, Universidad de Atacama, Copayapu 485, Copiap\'o, Chile, Email: \href{mailto:}{jose.fernandez@uda.cl and/or jfernandezt87@gmail.com}}
\altaffiltext{2}{Institut Utinam, CNRS-UMR 6213, Universit\'e Bourgogne-Franche-Compt\'e, OSU THETA Franche-Compt\'e, Observatoire de Besan\c{c}on, BP 1615, 251010 Besan\c{c}on Cedex, France, Email: \href{mailto:}{jfernandez@obs-besancon.fr}}
\altaffiltext{3}{Department of Physics and JINA Center for the Evolution of the Elements, University of Notre Dame, Notre Dame, IN 46556, USA}
\altaffiltext{4}{Instituto de Astronom\'ia, Universidad Nacional Aut\'onoma de M\'exico, Apdo. Postal 70264, M\'exico D.F., 04510, M\'exico}
\altaffiltext{5}{Universidade Federal do Rio Grande do Sul, Instituto de F\'isica, Av. Bento Gon\c{c}alves 9500, Porto Alegre, RS, Brazil}
\altaffiltext{6}{Depto. de Cs. F\'isicas, Facultad de Ciencias Exactas, Universidad Andr\'es Bello, Av. Fern\'andez Concha 700, Las Condes, Santiago, Chile.}
\altaffiltext{7}{Millennium Institute of Astrophysics, Av. Vicuna Mackenna 4860, 782-0436, Santiago, Chile.}
\altaffiltext{8}{Vatican Observatory, V00120 Vatican City State, Italy.}
\altaffiltext{9}{School of Physics and Astronomy, Sun Yat-sen University, Zhuhai 519082, China}
\altaffiltext{10}{Universidade de S\~ao Paulo, IAG, Rua do Mat\~ao 1226, Cidade Universit\'aria, S\~ao Paulo 05508-900, Brazil}
\altaffiltext{11}{Departamento de Astronom\'\i a, Casilla 160-C, Universidad de Concepci\'on, Concepci\'on, Chile}

\begin{abstract}
We report the discovery of a unique collection of metal-poor giant-stars, that exhibit anomalously high levels of $^{28}$Si, clearly above typical Galactic levels. Our sample spans a narrow range of metallicities, peaking at $-1.07\pm 0.06$, and exhibit abundance ratios of [Si,Al/Fe] that are as extreme as those observed in Galactic globular clusters (GCs), and Mg is slightly less overabundant. In almost all the sources we used, the elemental abundances were re-determined from high-resolution spectra, which were re-analyzed assuming LTE. Thus, we compiled the main element families, namely the light elements (C, N), $\alpha-$elements (O, Mg, Si), iron-peak element (Fe), $\textit{s}-$process elements (Ce, Nd), and the light odd-Z element (Al). We also provide dynamical evidence that most of these stars lie on tight (inner)halo-like and retrograde orbits passing through the bulge. Such kinds of objects have been found in present-day halo GCs, providing the clearest chemical signature of past accretion events in the (inner) stellar halo of the Galaxy, formed possibly as the result of dissolved halo GCs. Their chemical composition is, in general, similar to that of typical GCs population, although several differences exist. 
\end{abstract}

\section{Introduction} 
\label{section1}

\begin{figure*}[t]
	\begin{center}
		\includegraphics[height = 9 cm]{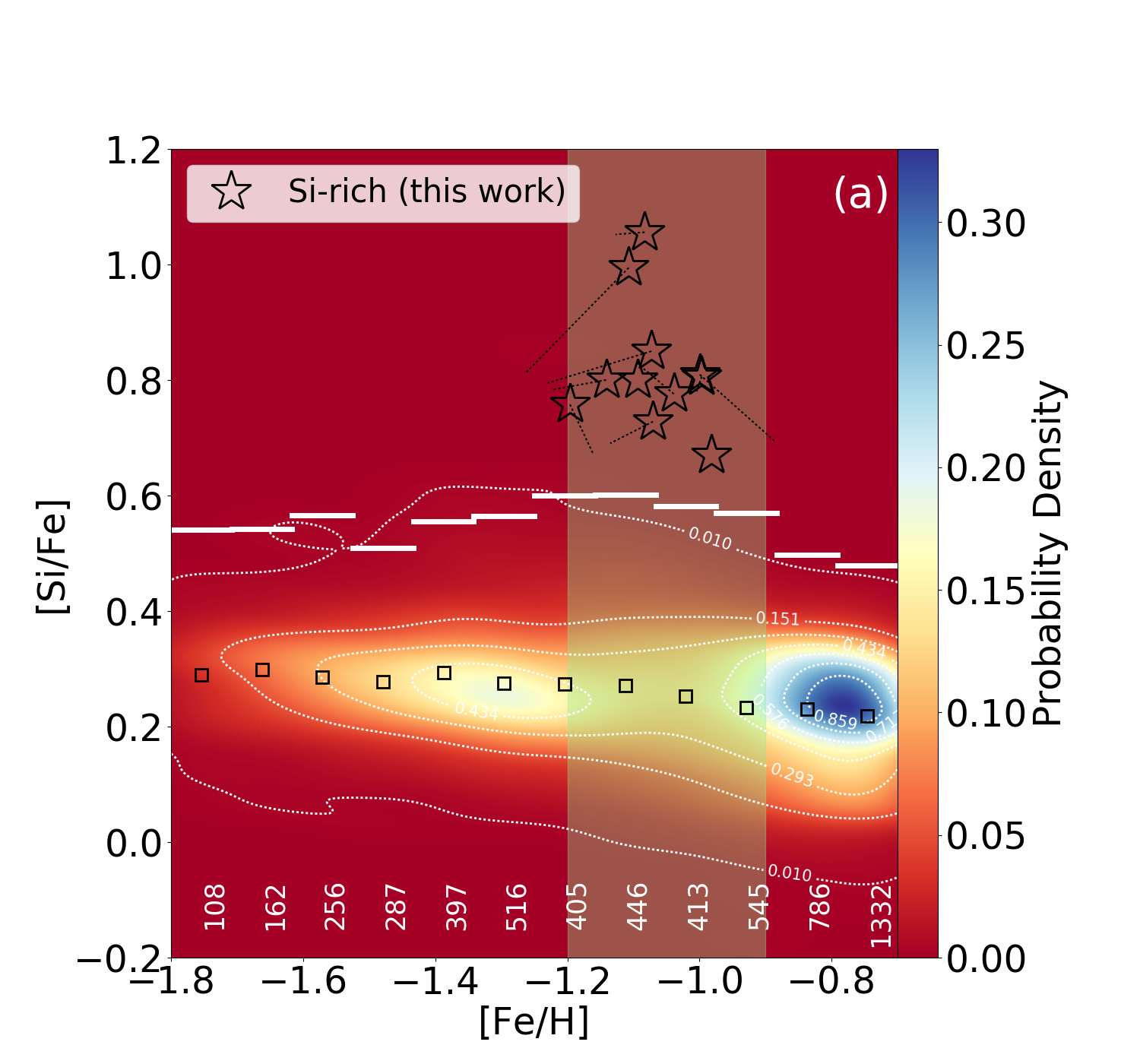}\includegraphics[height = 8.0 cm]{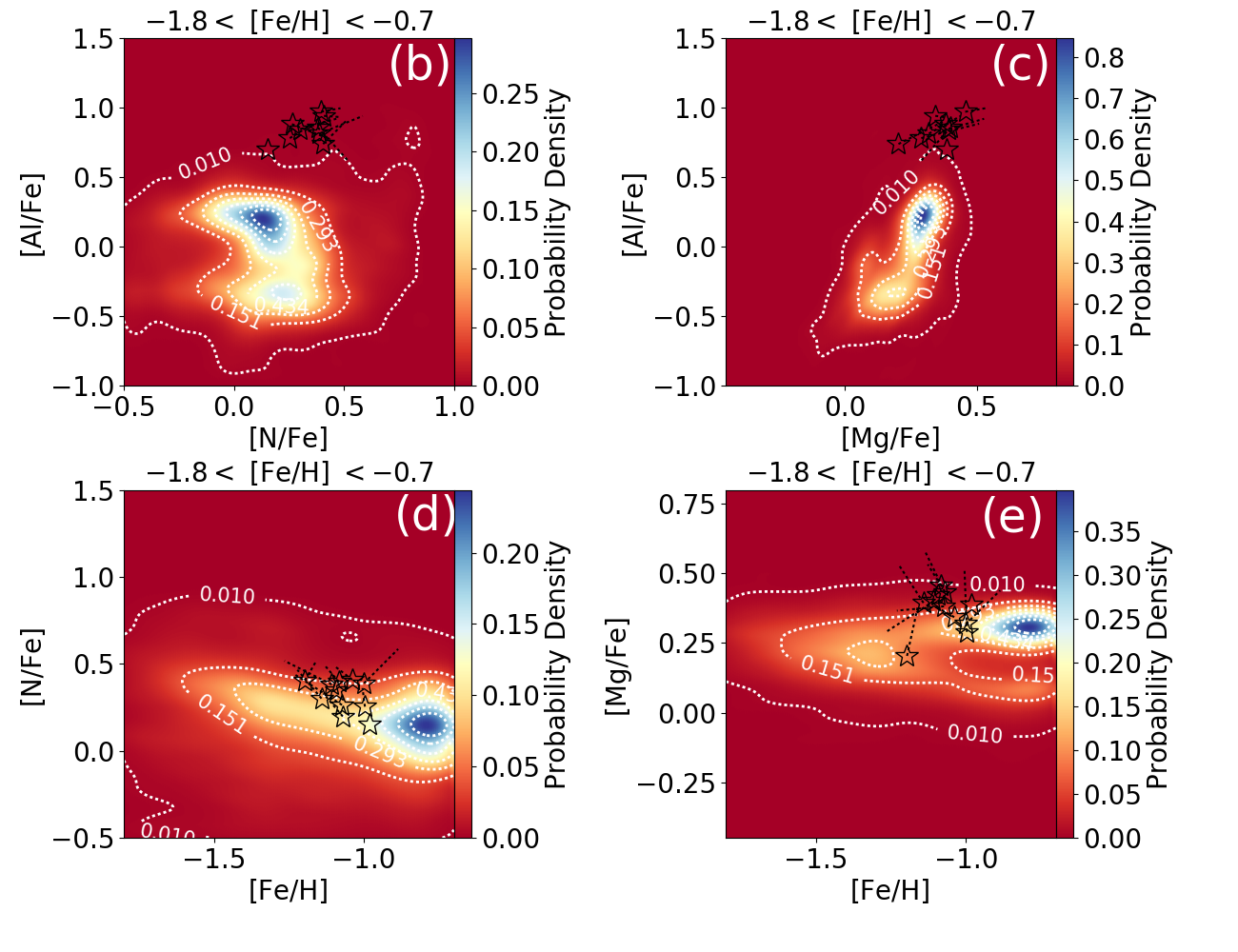}\\
		\caption{Kernel Density Estimation (KDE) of data displayed as a density colormap (panels a, b, c, d, and e), with a set of density contours as white dotted lines, whilst the colour indicates the probability density. The black unfilled star symbols refers to the abundance ratios ([X/Fe]$_{\rm spe}$) of the silicon-rich stars, whilst the black dotted line show the [X/Fe]$_{\rm pho}$ abundance ratios (see Table \ref{table1}). In (a) a shaded vertical area highlights the metallicity range ($-1.2 <$ [Fe/H] $< -0.9 $ dex) in which stars exhibit dramatic enhancement in silicon abundance, whilst the number of stars of each bin in the main body are shown at the center of the bin at the inner--bottom panel, and the black unfilled squares show the mean value of [Si/Fe] by bin, and the white line--segment indicate the selection criteria where [Si/Fe] $ > {\rm \langle [Si/Fe] \rangle} +3\sigma_{\rm [Si/Fe]}$. The error bars are estimates of the uncertainties ($\sigma_{total}  = \sqrt{\sigma^2_{[X/H], T_{\rm eff}}    + \sigma^2_{[X/H],{\rm log} g} + \sigma^2_{[X/H],\xi_t}  + \sigma^2_{mean}   }$) computed at the same way as \citet{Fernandez-Trincado2019b}. 
			}
		\label{fig1}
	\end{center}
\end{figure*}

The existence of chemical anomalies in field stars has been known for long, especially in the form of non-evolved or scarcely evolved stars on the Main Sequence (MS) and evolved Red Giant Branch (RGB) stars. While there is support in the literature that a small fraction, $\sim 1$--13\% of the present-day population of the stellar halo \citep[][]{Martell2016, Koch2019} of the Milky Way is filled of a unique collection of stars which are extremely common in clusters, and also characterised by distinctive chemical anomalies in a number of light elements, namely Al, N, Na and K with respect to Fe, and well above typical Galactic levels at a range of metallicities, to date a unifying explanation of the processes acting to produce such high levels of abundance anomalies or nucleosynthetic pathways in field stars remains elusive \citep{Fernandez-Trincado2016, Pereira2019a, Fernandez-Trincado2019a, Fernandez-Trincado2019b}. Hence, the more common assumption is that the majority of such unexplained objects that reside in the field which replicate or exceed the extreme abundance patterns seen in GCs are commonly referred to as tidally-dissipated remmants of GCs (unless they are part of a binary system), as they have similar properties to that of the present day GCs  \citep[][among other]{Martell2016, Schiavon2017a, Fernandez-Trincado2016, Fernandez-Trincado2017L, Fernandez-Trincado2019b, Fernandez-TrincadoOrbit, Koch2019}, with only a handful of exceptions which are often associated to other stellar associations like dwarf galaxies \citep[e.g.,][]{Hasselquist2019}.

Among the variety of objects with exotic chemical composition that populate Galactic GCs, stars with anomalously high levels of  [Si/Fe] ratios ($\gtrsim+0.6$, which we henceforth refer to as silicon-rich stars) among metal-poor stars are surely among those still presenting many riddles to astronomers. The Si overabundance of GCs stars can be produced, under some assumptions, i.e., the Mg-Si anti-correlation found in several GCs \citep{Carretta2019} is of particular importance, because it pinpoint the "leakage" mechanism from Mg-Al cycle on $^{28}$Si as the major source of the phenomenon, i.e., when the $^{27}$Al(p,$\gamma$)$^{28}$Si reaction takes over $^{27}$Al(p,$\alpha$)$^{24}$Mg a certain amount of $^{28}$Si is produced by p-captures \citep{Karakas2003}. However, temperatures in excess upwards of $>$ 100MK are required to activate the relevant reactions attained in the polluters, which thus provide the overproduction of [Si/Fe] ratios \citep[see e.g.,][]{Arnould1999}, and apparently possible only among stars in a subset of GCs stars \citep{Masseron2019}. These silicon-rich stars apparently tend to form at all metallicities \citep{Carretta2019}, so it is probable that the polluting source is extant at all metallicities.

\begin{figure}[t]
	\begin{center}
		\includegraphics[height = 9.5 cm]{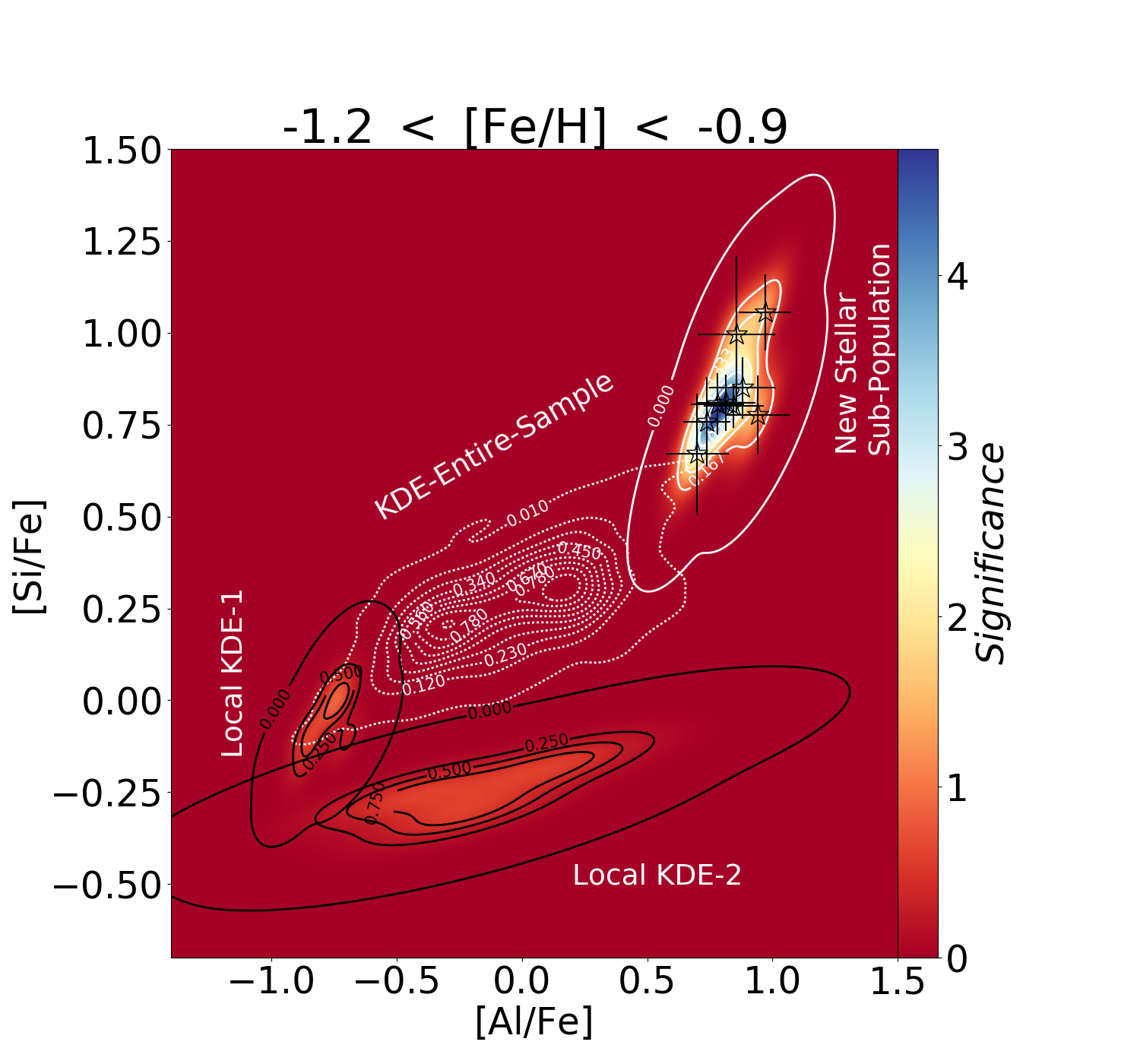}\\
		\caption{Kernel Density Estimation (KDE) models in the [Si, Al/Fe] plane for the entire sample (white dashdot contour) in the metallicity range $-1.2 < $ [Fe/H] $< -0.9$ compared to the KDE models of two arbitrary local volume that contain some data points (black line contours), and the KDE of the New Stellar Sub-Population (white line contour). The colour indicates the significance of each local volume as compared to the back ground level.}
		\label{fig22}
	\end{center}
\end{figure}

Thus, the task to identify of such kind of stars beyond GCs environments may reveal any chemical peculiarity and also allow to constraint the search for stellar tidal debris of defunct GCs in the (inner) stellar halo of the Milky Way, implying that $^{28}$Si would be an important chemical specie for a genetic link to GCs stars, thus playing a role in determining whether light element anomalies arise from nuclear processing within evolved stars born within GCs environments or tied to a certain epoch in the Milky Way's evolution. It is also in line with previous suggestions that the (inner) stellar halo may have formed a non-negligible fraction of its mass \textit{ex situ}, according to which these stars originate from the dissolution of a population of accreted GCs \citep{Martell2016}. 

So far, stars with significantly enhanced aluminum paired with silicon enrichment have thus far only been identified in a few GCs environments \citep{Masseron2019}. The origin of such signs remain unknown, as does the reason for its apparent exlusivity to GCs environments. In this Letter we report on the serendipitous discovery by the Apache Point Observatory Galactic Evolution Experiment \citep[APOGEE-2, ][]{Majewski2017} of a unique collection of mildly metal-poor field giant-stars, which exhibit Silicon overabundances ([Si/Fe]$\gsim+0.6$), significantly higher than typical Galactic abundances of  [Si/Fe]$<+0.5$, which we hypothesized that all silicon-rich stars were formed in GCs environments and were later lost to the field when the stellar clusters were ultimately disrupted and that also contributed to the (inner) stellar halo buildup.

\floattable
	\begin{deluxetable*}{llccccccccccccc}
		\center
		\setlength{\tabcolsep}{0.7mm}  
		\tabletypesize{\small}
		\tablecolumns{17}
		\rotate
		\tablecaption{Renormalised unit weight error (RUWE) for 8 out of 11 Silicon-rich stars astrometrically well-behaved, and good inferred distances using \texttt{StarHorse} (\texttt{gdr2\_contrib.starhorse}). In this table, we consider only the subset of stars in the sample that respects the flags recommended in \citet{Anders2019} which corresponds to the most robust distances, i.e., \texttt{SH\_GAIAFLAG}$=$ '000' and \texttt{SH\_OUTFLAG}$=$ '00000' }
		\tablehead{
\colhead{APOGEE-2 ID   }                            &  
\colhead{NgAl                }                            &  
\colhead{chi2AL             }                            &  
\colhead{Np                   }                             & 
\colhead{Gmag              }                             & 
\colhead{BPmag             }                             &  
\colhead{RPmag             }                             & 
\colhead{RUWE               }                             & 
\colhead{\texttt{SH\_GAIAFLAG}   }              & 
\colhead{\texttt{SH\_OUTFLAG}    }              & 
\colhead{d$_{\odot}$                }              & 
\colhead{RV                               }              & 
\colhead{$\mu_{\alpha}\cos(\delta)$          }  & 
\colhead{$\mu_{\delta}$           } &
\colhead{Galactic Orbit} \\
\colhead{}		    &   
\colhead{}		    & 
\colhead{}		    &    
\colhead{}		    &   
\colhead{}		    &  
\colhead{}		    &    
\colhead{}		    &        
\colhead{}		    & 
\colhead{}		    & 
\colhead{}		    & 
\colhead{kpc                    } & 
\colhead{km s$^{-1}$      } & 
\colhead{mas yr$^{-1}$   } & 
\colhead{mas yr$^{-1}$   } &
\colhead{  }\\
\colhead{}		    &   
\colhead{}		    & 
\colhead{}		    &    
\colhead{}		    &   
\colhead{}		    &  
\colhead{}		    &    
\colhead{}		    &        
\colhead{}		    & 
\colhead{}		    & 
\colhead{}		    & 
\colhead{\texttt{StarHorse}     } & 
\colhead{\texttt{APOGEE-2}    } & 
\colhead{\texttt{Gaia DR2}      } & 
\colhead{\texttt{Gaia DR2}      } &
\colhead{\texttt{GravPot16}     }
		}
		\startdata
			2M13314691$+$2804210 &    560  &  694.07  &  15  & 12.45   & 13.21  &  11.61   & 1.02       & '000'       &  '00000' & 6.14$\pm$1.94   &  $-1.72\pm0.12$   &  $-7.32\pm0.08$    & $-6.54\pm0.05$ & $\checkmark$ \\
			2M15153684$+$3501283 &  292  &  178.51  &  18  & 13.88   & 14.52  &  13.13   & 0.99       & '000'       &  '00000' & 11.0$\pm$3.66     &  $-308.45\pm0.12$   &  $-4.14\pm0.04$      & $-7.74\pm0.05$   & $\checkmark$\\
			2M16092248$+$2449223 &    476  &  753.02  &  18  & 12.41   & 13.25  &  11.53   & 1.09       & '000'       &  '00000' & 6.34$\pm$1.33   &  $-92.29\pm0.22$   &  $-9.28\pm0.05$      & $-1.51\pm0.05$  & $\checkmark$\\
			2M16300791$+$2537503 &   354  &  294.97  &  17  & 13.03   & 13.58  &  12.34   & 0.96       & '000'       &  '00000' &  4.89$\pm$0.57   & $-255.29\pm0.02$   &  $-3.02\pm0.04$      & $-9.39\pm0.04$  & $\checkmark$ \\
			2M16482145$-$1930487 &   301  &  276.21  &  12  & 14.13   & 15.04  &  13.18   & 1.00       & '000'       &  '00000' & 7.88$\pm$1.69     &  $65.99\pm0.03$   &  $-0.11\pm0.07$      & $-10.05\pm0.05$ & $\checkmark$ \\
			2M17155274$+$2907368 &   279  &  203.76  &  17  & 13.77   & 14.44  &  12.99   & 1.09       & '000'       &  '00000'  & 7.69$\pm$1.48    &  $-81.52\pm0.03$   &  $-4.11\pm0.04$      & $-1.87\pm0.05$  & $\checkmark$ \\
			2M17214096$+$4246147 &    185  &  194.90  &  16  & 12.83   & 13.48  &  12.07   & 0.91       & '000'       &  '00000' & 5.97$\pm$1.40    &  $-298.07\pm0.05$   &  $-5.59\pm0.05$      & $-3.97\pm0.06$  & $\checkmark$\\
			2M23341347$+$4836321 &    400  &  313.58  &  18  & 13.88   & 14.75  &  12.97   & 0.98        & '000'       &  '00000' & 11.04$\pm$3.01 & $-168.33\pm0.15$   &  $2.28\pm0.05$      & $-0.43\pm0.05$ & $\checkmark$ \\
		\hline
		\enddata
		\tablecomments{
				NgAl $=$ \texttt{astrometric\_n\_good\_obs\_al} -- chi2AL $=$ \texttt{astrometric\_chi2\_al} -- Np $=$ \texttt{VISIBILITY\_PERIODS\_USED} -- Gmag $=$ \texttt{phot\_g\_mean\_mag} -- BPmag $=$ \texttt{phot\_bp\_mean\_mag} -- RPmag $=$ \texttt{phot\_rp\_mean\_mag}. For each PM component we have accounted for the systematic uncertainty on the order of 0.035 mas yr$^{-1}$ \citep{Lindegren2018}.
		}					
		\label{table2}
	\end{deluxetable*}
	
\section{Discovery of a New Stellar Sub-population}
\label{sec:sec2}

We report the discovery of an unexpected and exceptional enhancement of [Si/Fe] abundance ratios among mildly metal-poor giant-stars ([Fe/H]$< -0.7$) in the Galactic field, which have been identified from the $\sim$ 270,000 APOGEE-2 giant-stars \citep[14th data release of SDSS, ][]{Abolfathi2018}, that exhibit significantly enhanced [Si/Fe] and [Al/Fe] abundance ratios. Since we are primarly interested in abundance anomalies of metal-poor stars, our focus in this work is on giant-stars with metallicities [Fe/H] $<-0.7$\footnote{By requiring metallicity more metal-poor than $-0.7$ we minimize the presence of giants in the thin and thick disk, and by imposing a lower limit on metallicity of $-$1.8, allow for the inclusion of stars with reliable carbon and nitrogen abundances \citep[see][and reference therein]{Fernandez-Trincado2019b}}. We selected a sample of giant-stars, adopting conservative cuts on the columns of the APOGEE-2 catalogue in the following way: \textit{(i)}  S/N $>$ 70 pixel$^{-1}$; \textit{(ii)} 3500 K $<$T$_{\rm eff}$ $<$ 6000 K; \textit{(iii)} log \textit{g} $<$ 3.6; and (\textit{iv}) \texttt{ASPCAPFLAG}\footnote{This cut ensure that there were no major flagged issues, i.e., low S/N, poor synthetic spectral fit, stellar parameters near grid boundaries, among other.} $=$ 0. Following the conventions in the literature \citep[see e.g.,][]{Schiavon2017a}, we require non-enhanced carbon stars with [C/Fe] $<+0.15$, because such stars are commonly found in globular clusters, and the surface abundance of stars enhanced in carbon ([C/Fe]$>+0.15$) may have been modified by mass transfer from an AGB companions \citep[e.g.,][]{Lucatello2006}. The final sample so selected amounts to a total of 5,653 stars. The method in deriving atmospheric parameters and abundances is identical to that of \citet{Fernandez-Trincado2019b} adopting the Brussels Automatic Stellar Parameter (\texttt{BACCHUS}) code \citep{BACCHUS}. The results are listed in Table \ref{table1}. 

\clearpage
\floattable
\begin{deluxetable*}{ccccccccccccccccc}
	\center
	\setlength{\tabcolsep}{0.7mm}  
	\tabletypesize{\tiny}
	\tablecolumns{17}
	\rotate
	\tablecaption{\textit{Rows 1--13:} Adopted atmospheric parameters of our target stars, frequency of observation per object (N$_{\rm visits}$), signal-to-noise (S/N), radial velocity (\textit{RV}) and radial velocity scatter information ($RV_{SCATTER}$). The average value of the orbital elements (pericentric and apocentric radii, the maximum distance the orbit reaches above/below the Galactic plane, and the eccentricity was found for one million realizations, with uncertainty ranges given by the 16th (subscript) and 84th (superscript) percentile values. \textit{Rows 14--37:} Mean elemental abundances derived of our target stars using the "abund" module in \texttt{BACCHUS} code, adopting the atmospheric parameters from \texttt{ASPCAP}/APOGEE-2 (abundances labeled as [Fe/H]sp and [X/Fe]sp) and atmospheric parameters from photometry and isochrones (abundances labeled as [Fe/H]pho and [X/Fe]pho). The Solar reference abundances are from (Asplund2005). The \texttt{BACCHUS} pipeline was used to derive the broadening parameters, metallicity, and chemical abundances.}
	\tablehead{
					\colhead{APOGEE-2 ID}             &  \colhead{[M/H]} &   \colhead{$J_{2MASS}-K_{s,2MASS}$}   &   \colhead{E(B-V)}  & \colhead{T$^{pho}_{\rm eff}$} & \colhead{log$^{\rm iso}$} & \colhead{Teff$^{\ddagger}_{ASPCAP}$} & \colhead{logg$^{\ddagger}_{ASPCAP}$} & \colhead{N$_{\rm visits}$} & \colhead{S/N} & \colhead{\textit{RV}} & \colhead{$RV_{SCATTER}$ } & \colhead{$\langle$r$_{\rm peri} \rangle$} & \colhead{$\langle$r$_{\rm apo}\rangle$} & \colhead{$\langle$Z$_{\rm max} \rangle$}  & \colhead{$\langle e \rangle$} & \colhead{Orbit} \\
					\colhead{}			& \colhead{dex} &  \colhead{mag} & \colhead{} & \colhead{K} & \colhead{dex} &  \colhead{K} & \colhead{dex} & \colhead{} & \colhead{pixel$^{-1}$} & \colhead{km s$^{-1}$} & \colhead{km s$^{-1}$} & \colhead{kpc} & \colhead{kpc} & \colhead{kpc} & \colhead{} & \colhead{}
	}
	\startdata
			2M13314691$+$2804210$^{\dagger}$ &    $-$1.08 &    0.907 &    0.009 &    4048.1 &     0.646 &     4114.6 &     1.381   &   4       &  248   &    $-$1.72 &   0.12    &  0.49$^{0.56}_{0.22}$ &  10.25$^{10.70}_{9.89}$ & 6.33$^{7.53}_{6.15}$ & 0.91$^{0.96}_{0.89}$  & Retrograde \\
			2M15153684$+$3501283 &    $-$0.99 &    0.77  &    0.017 &    4401.3 &     1.341 &     4379.2 &     1.808   &   3       &  115   &  $-$308.45 &   0.12    &  1.84$^{1.88}_{1.78}$ &  20.13$^{25.59}_{17.75}$ & 15.81$^{23.85}_{13.55}$ & 0.84$^{0.86}_{0.81}$  & Retrograde \\
			2M16092248$+$2449223$^{\dagger}$ &    $-$1.14 &    0.971 &    0.070 &    3990.5 &     0.539 &     4095.6 &     1.237   &   5       &  364   &   $-$92.30 &   0.22   &  0.11$^{0.96}_{0.05}$ &  9.01$^{9.25}_{8.45}$ & 6.24$^{7.62}_{5.39}$ & 0.98$^{0.99}_{0.81}$  & Retrograde  \\
			2M16300791$+$2537503 &    $-$1.09 &    0.67  &    0.035 &    4717.5 &     1.851 &     4785.8 &     2.169   &   2       &  143   &  $-$255.29 &   0.02   &  0.61$^{0.65}_{0.53}$ &  7.65$^{7.72}_{7.46}$ & 6.81$^{6.92}_{6.78}$ & 0.85$^{0.87}_{0.84}$  & Retrograde  \\
			2M16482145$-$1930487 &    $-$0.96 &    1.013 &    0.520 &    4638.1 &     1.767 &     4377.4 &     1.741   &   3       &   97   &      65.98 &   0.01    &  0.15$^{0.35}_{0.02}$ &  4.96$^{5.48}_{4.29}$ & 2.64$^{3.39}_{2.29}$ & 0.93$^{0.99}_{0.88}$  & Prograde \\
			2M17155274$+$2907368 &    $-$1.07 &    0.839 &    0.058 &    4287.6 &     1.076 &     4313.9 &     1.601   &   2       &   94   &   $-$81.42 &   0.00    &  0.76$^{0.82}_{0.65}$ &  9.19$^{9.68}_{8.93}$ & 6.98$^{8.63}_{6.05}$ & 0.85$^{0.87}_{0.83}$   & Prograde \\
			2M17214096$+$4246147 &    $-$1.11 &    0.777 &    0.020 &    4385.1 &     1.249 &     4439.0 &     1.614   &   5       &  181   &  $-$298.07 &   0.05   &  1.43$^{1.46}_{1.42}$ &  9.34$^{10.16}_{9.06}$ & 4.03$^{4.48}_{3.84}$ & 0.73$^{0.75}_{0.72}$  & Retrograde  \\
			2M17502038$-$2805411 &    $-$1.09 &    3.492 &    ...   &    ...    &     ...   &     3775.0 &     0.487   &   2       &  101   &   $-$53.06 &   0.13     &  ... &  ... & ... &  ... &  ... \\
			2M18070782$-$1517393$^{\dagger}$ &    $-$1.07 &    1.66  &    ...   &    ...    &     ...   &     3924.7 &     1.126   &   3       &  151   &  $-$370.42 &   0.02    &  ... &  ... & ... &  ... &  ... \\
			2M19281906$+$4915086 &    $-$1.22 &    0.79  &    0.086 &    4460.6 &     1.303 &     4309.0 &     1.715   &   3       &  422   &  $-$302.89 &   0.18    &  ... &  ... & ... &  ... &  ... \\
			2M23341347$+$4836321 &    $-$1.17 &    1.027 &    0.196 &    4048.9 &     0.566 &     4073.1 &     1.227   &   3       &  205   &  $-$168.33 &   0.15     &  0.29$^{0.54}_{0.11}$ &  17.76$^{20.21}_{17.29}$ & 3.20$^{4.69}_{2.98}$ & 0.97$^{0.99}_{0.94}$ & Prograde \\
			\hline
			\hline
						           & [Fe/H]sp   & [C/Fe]sp   & [N/Fe]sp  &    [O/Fe]sp  & [Mg/Fe]sp &  [Al/Fe]sp   &  [Si/Fe]sp  &        [Ce/Fe]sp  &  [Nd/Fe]sp         \\
						\hline
						2M13314691$+$2804210 & $-$1.07    &   0.02    &  0.26    &    0.43    &   0.38    &   0.88    &   0.85     &  0.73     & ...    \\
						2M15153684$+$3501283 & $-$0.99    &   0.12    &  0.25    &    0.43    &   0.29    &   0.78    &   0.81     &  0.65     & ...    \\
						2M16092248$+$2449223 & $-$1.11    &$-$0.17    &  0.38    &    0.43    &   0.40    &   0.86    &   0.99     &  0.72     & 0.77   \\
						2M16300791$+$2537503 & $-$1.04    &   0.01    &  0.41    &    0.50    &   0.34    &   0.94    &   0.78     &  0.33     & ...    \\
						2M16482145$-$1930487 & $-$0.99    &   0.12    &  0.39    &    0.50    &   0.32    &   0.81    &   0.81     &  ...      & 0.89   \\
						2M17155274$+$2907368 & $-$1.07    &   0.01    &  0.20    &    0.41    &   0.43    &   ...     &   0.73     &  0.30     & ...    \\
						2M17214096$+$4246147 & $-$1.08    &$-$0.16    &  0.40    &    0.44    &   0.46    &   0.97    &   1.06     &  0.76     & ...    \\
						2M17502038$-$2805411 & $-$1.09    &$-$0.14    &  0.35    &    0.42    &   0.42    &   ...     &   0.80     &  0.45     & ...    \\
						2M18070782$-$1517393 & $-$0.98    &   0.10    &  0.15    &    0.33    &   0.39    &   0.70    &   0.67     &  0.70     & ...    \\
						2M19281906$+$4915086 & $-$1.20    &$-$0.21    &  0.41    &    0.36    &   0.20    &   0.74    &   0.76     &  ...      & ...    \\
						2M23341347$+$4836321 & $-$1.14    &$-$0.04    &  0.30    &    0.40    &   0.39    &   0.85    &   0.80     &  0.70     & ...    \\											
						\hline
						\hline
						&   [Fe/H]pho    &  [C/Fe]pho   & [N/Fe]pho  &   [O/Fe]pho  &   [Mg/Fe]pho  &    [Al/Fe]pho &  [Si/Fe]pho   &   [Ce/Fe]pho &  [Nd/Fe]pho     \\		
						\hline
						2M13314691$+$2804210 & $-$1.23    &$-$0.09    &  0.44    &    0.44    &   0.37    &   0.77    &   0.79     &  0.41     &  ...    \\
						2M15153684$+$3501283 & $-$1.00    &$-$0.01    &  0.36    &    0.50    &   0.51    &   0.88    &   0.78     &  0.38     &  ...    \\
						2M16092248$+$2449223 & $-$1.26    &$-$0.26    &  0.52    &    0.35    &   0.29    &   0.62    &   0.81     &  0.41     &  0.41   \\
						2M16300791$+$2537503 & $-$1.09    &$-$0.04    &  0.48    &    0.48    &   0.38    &   0.93    &   0.83     &  0.39     &  ...    \\
						2M16482145$-$1930487 & $-$0.89    &   0.06    &  0.59    &    0.81    &   0.44    &   0.94    &   0.69     &  ...      &  0.81   \\
						2M17155274$+$2907368 & $-$1.14    &$-$0.08    &  0.33    &    0.44    &   0.58    &   ...     &   0.69     &  0.32     &  ...    \\
						2M17214096$+$4246147 & $-$1.13    &$-$0.24    &  0.48    &    0.42    &   0.53    &   0.99    &   1.05     &  0.52     &  ...    \\
						2M17502038$-$2805411 &  ...       &   ...     &  ...     &    ...     &   ...     &   ...     &   ...      &  ...      &  ...    \\
						2M18070782$-$1517393 &  ...       &   ...     &  ...     &    ...     &   ...     &   ...     &   ...      &  ...      &  ...    \\
						2M19281906$+$4915086 & $-$1.16    &$-$0.28    &  0.51    &    0.63    &   0.38    &   0.90    &   0.67     &  ...      &  ...    \\
						2M23341347$+$4836321 & $-$1.22    &$-$0.17    &  0.47    &    0.44    &   0.53    &   0.92    &   0.78     &  0.36     &  ...   \\ 	
						\hline
				\enddata
				\tablecomments{$^{\ddagger}$Uncalibrated ASPCAP parameters are listed in columns 7 and 8. 
					}					
\label{table1}
\end{deluxetable*}

A new stellar sub-population was identified for searching for outliers in the [Si/Fe]--[Fe/H] abundance plane. We determined the boundary between these outliers and the Milky Way stars by identifying the trough in the [Si/Fe] distribution over 12 metallicity bins and performing a single gaussian component fit. We label all stars with [Si/Fe] abundance more than $3\sigma_{\rm [Si/Fe]}$ above the median [Si/Fe] abundance value of the corresponding iron bin as "silicon-rich" (see Figure \ref{fig1}). The bins were chosen to ensure that at least $\sim$100 stars were in each bin. This returns 36 stars that appear silicon-rich relative to the full data set. We then eliminate three known GCs stars from, leaving 33 silicon-rich candidates.

 In order to confidently state chemical tagging results for this data set, we also require enhanced aluminum ([Al/Fe]$\gsim+0.5$), as a lower aluminum abundance is not by itself a sufficient indicator of a GC-like abundance pattern in the Galactic field. If high levels of Al enrichments are found, this would imply that such Si-Al-rich giant field stars and \textit{second-generation} GCs stars could potentially be the same kind of stellar objedts, with similar nucleosynthetic histories. This leaves us with 11 likely Si-Al-rich red-giants with high-quality spectra and reliable parameters and abundances. It is to note that for the newly identified silicon-rich stars, there are no nearby GCs showing direct association, based on the chemical/dynamical properties. To further assess the statistical significance of the newly identified stellar sub-population, we ran a Kernel Density Estimation (KDE) over this sample, and compared them with the tail of the main Milky Way locus (silicon-normal stars), i.e., with some arbitrary local volume that contains some data points as illustrated in Figure \ref{fig22}. Finally, in Figure \ref{fig22}, we plot the KDE models over our data to demonstrate that there is a true low density valley separating the silicon-normal population from the silicon-rich population, it gives us an idea of the significance of our finding, which exceed the background level by a factor of $\sim$4, which is also significant as compared to the two arbitrary local volume of the entire sample. A set of contour lines is set to visual aid. From Figure \ref{fig22}, we can see how the local "silicon-rich" sample occupies a separate and significant locus of stars beyond [Si/Fe]$>+0.5$, which is not on the main bulk of the KDE of the entire sample.

\begin{figure}[t]
	\begin{center}
		\includegraphics[height = 13 cm]{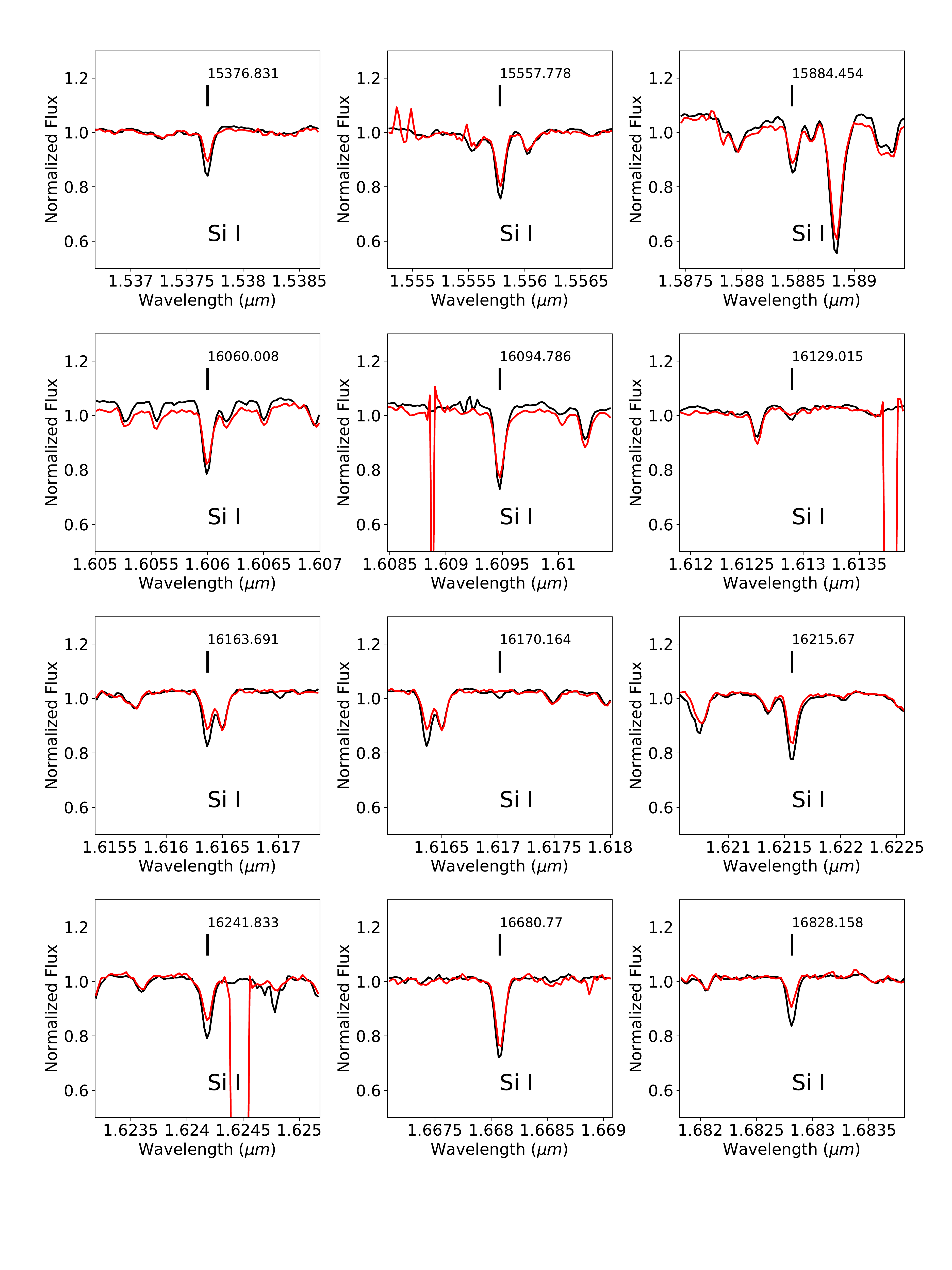}\\
		\caption{
			{Comparison between the APOGEE-2 spectrum of a normal (red line, star: 2M14424814$+$4653219, T$_{\rm eff} = 4346$ K, log \textit{g}$=$1.66, [Fe/H]$=-$1.19, [Si/Fe]$=+$0.25) and a silicon-rich giant star (black line, star: 2M19281906$+$4915086, T$_{\rm eff} = 4309$ K, log \textit{g}$=$1.71, [Fe/H]$=-$1.20, [Si/Fe]$=+$0.65) around the Si I lines, with similar stellar parameters.}
		}
		\label{fig2}
	\end{center}
\end{figure}

A comparison is presented in Figure \ref{fig2}, where the spectra of a silicon-rich and a silicon-normal star are shown in the relevant wavelength range containing the Si I lines, indicated by vertical tick marks. The silicon-rich star has remarkably stronger Si I lines which, in view of the similarity between the two stars in all the other relevant parameters, can only mean that it has a much higher silicon abundance. Both \texttt{ASPCAP} \citep{ASPCAP} and our line-by-line manual analysis with \texttt{BACCHUS} tell us that [Si/Fe] in the silicon-rich star is higher than the silicon-normal star by $\sim3.73\sigma$. In doing chemical tagging in the disk (thick), we have moved past just asking whether stars could have formed in GCs, and now we are looking for ways to understand several different types of chemically anomalous stars.

\section{RESULTS AND DISCUSSION}

The Si overabundances reported here are significantly above the typical value measured in the Milky Way, and striking similar those observed in Galactic GCs environments. 

\subsection{Chemical evidence for a new stellar sub-population in the inner stellar halo of the Milky Way} 

Figure \ref{fig1} (\textit{top} panel) tantalizingly suggest that a new sub-population of stars chemically differentiated by their highest [Si/Fe] abundance ratios has been identified in the Galaxy, clumped at values [Si/Fe]$\sim +0.82\pm 0.11$ (which we henceforth refer to as silicon-rich stars), and clearly separated from the main body of silicon-normal stars for [Si/Fe]$<+0.5$. Regarding their metallicity distribution, the silicon-rich stars range between $-1.2\lsim$ [Fe/H] $\lsim-0.9$, with a mean value of [Fe/H] $\pm \sigma_{\rm [Fe/H]}$ $= -1.07 \pm 0.06$, suggesting in fact a possible association with the thick disk or halo. If we make a simplistic assumption that Milky Way stars in that metallicity range can be modeled as a Gaussian with a mean of [Si/Fe] $=0.26$ and $\sigma_{\rm [Si/Fe]} = 0.11$ (over 1458 stars in that iron bin), the probability of drawing a silicon-rich star ([Si/Fe]$>0.67$) star at the same metallicity bin is $<$0.1 per cent ($3.73\sigma$). The probability of drawing a star of that metallicity with [Si/Fe]$>0.6$ from the Galaxy population is small. As such, the next is devoted to exploring these object in more detail. 

Figure \ref{fig1}(a) show the trend of [Si/Fe] with metallicity for accreted halo stars, canonical halo and thick disk stars with [Si/Fe]$< +0.6$ dex (silicon-normal stars), which  appear to be very mixed according to \citet[][]{Hawkins2015}. Hence, the discovery of such silicon-rich stars, suggest that $^{28}$Si might follow different nucleosynthesis pathway(s), and reveals a well-defined dichotomy between field giants and other Galactic environments (like GCs stars), and it is possible that such objects may represent a population for stars evaporated from GCs.

From Figure \ref{fig1}(e,f) we also note that silicon-rich stars exhibit slightly enriched levels of N and Mg than most Milky Way stars at the same metallicities, at least as far as observed dispersion are concerned, i.e., $\sigma_{\rm N} = 0.09$ dex and $\sigma_{\rm Mg} = 0.07$ dex, respectively, indicating a distinct formation history. In addition, there are chemistry similarities in [Mg/Fe] and [N/Fe] ratios between the silicon-rich stars and the population of stars in the innermost regions of the Galactic halo. Thus, N and Mg affirm that the silicon-rich could be a sub-population of the same population as the high-Mg halo population \citep[see, e.g.,][]{Hayes2018}.

We emphasize that the enhanced abundances of [Si/Fe] in these objects is not only caused by the natural [Si/Fe] variations observed for field stars. To our knowledge, the only Galactic environments that contains a sufficient fraction of stars with a sufficiently enrichement in [Si/Fe] paired by the modest enrichment in [Al/Fe] are Galactic GCs \citep[][]{Masseron2019} at similar metalliticity. We conclude that it is difficult to explain high [Si/Fe] values of these sources with the typical history of silicon enrichment of the Galaxy.

\textit{Galactic orbits:} We computed the more probable Galactic orbit for 8/11 silicon-rich stars. For this, we combine precise proper motions from Gaia DR2 \citep{Lindegren2018}, radial velocity from APOGEE-2 \citep{Nidever2015} and the inferred distances using \texttt{StarHorse} code \citep{Queiroz2018}, which include additionally the cuts (\texttt{SH\_GAIAFLAG} $=$ 
'000' and \texttt{SH\_OUTFLAG} $=$ '00000') described in \citet{Anders2019}, as input data in the new state-of-the-art orbital integration package \texttt{GravPot16}\footnote{\url{https://gravpot.utinam.cnrs.fr}}. Table \ref{table2} list the renormalised unit weight error (RUWE) and available distance for these sources from the \texttt{gdr2\_contrib.starhorse} table, where the typical RUWE is less than or equal to 1.4, indicating that these sources are 	astrometrically well-behaved. 
	
\begin{figure}[t]
	\begin{center}
		\includegraphics[height = 18 cm]{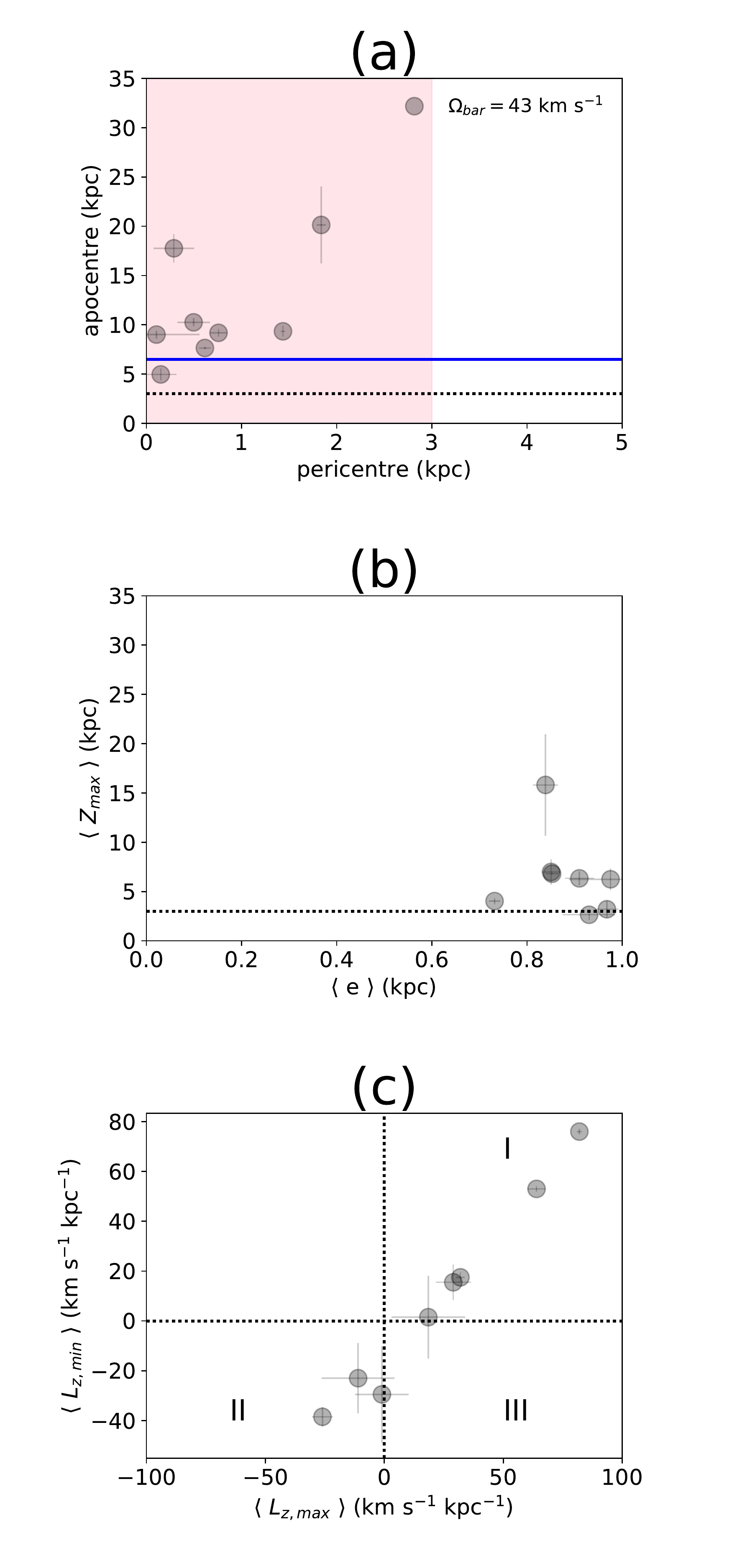}\\
		\caption{Orbital parameters of the silicon-rich stars. In panel (a), the shaded region and the black dotted line indicates the radius (3 kpc; \citet{Barbuy2018}) of the Milky Way bulge, whilst the blue line indicate the location of the bar's corotation radius (CR $\sim$ 6.5 kpc), a star below the black dotted line would have a bulge-like orbit. In panel (b), the black dotted line represent the edge $Z_{max}$ of the thick disk ($\sim$3 kpc, \citet{Carollo2010}). In panel (c), the black dotted lines divide the regions with prograde orbits (region \textit{II}) with respect to the direction of the Galactic rotation, retrograde orbits (\textit{region I}), and stars that have prograde-retrograde orbits at the same time (\textit{region III}). The error bars show the uncertainty in the computed orbital parameters.}
		\label{fig4}
	\end{center}
\end{figure}

	 Since the true Galactic potential is not accurately known in the inner Galaxy ($< 0.5$ kpc), the results of our simulations may depend significantly on the assumed model in the inner Galaxy, especially for those stars with orbital incursions closer to the Galactic centre, assuming a $\Omega_{\rm bar} = 43$ km s$^{-1}$ \citep{Bovy2019}. The silicon-rich stars are found to have radial orbits, with pericentre values less than 2 kpc, apocentre values ranging between 4--20 kpc, eccentricities larger than 0.5, and maximum vertical excursion from the Galactic plane ranging between 1.8--16 kpc. In Figure \ref{fig4} and Table \ref{table1} we give for each silicon-rich star some orbital parameters; we refer the reader to \citet{Fernandez-Trincado2019b} to details regarding to all the parameters employed in our Galactic model. We find that these are not stars that lives in the inner Galaxy, and therefore those identified toward the bulge region are likely halo intruders into the Galactic bulge. Most of the silicon-rich stars exhibit retrograde orbits, and have halo-like orbits, characterised by high eccentricities ($e>0.5$). It is important to note that the large spread in the apocenter distance, in addition of stars having a prograde or a retrograde sense (see Figure \ref{fig4}c) with respect to the rotation of the bar, suggest that they cannot have a common origin (unless they were released from their parent at a very different time). In this sense, the predict orbits would suggest that most of these stars were likely formed during the very early stages of the evolution of the Milky Way, in a similar way as Galactic GCs. These dynamical properties accompanied by the large enrichment in Si, as well as the modest enhancement in Al, and the no evidence for an intrinsic Fe abundance spread ($< 0.06$ dex), reveal that these objects were likely born from the same molecular cloud and later accreted during the initial phases of Galaxy assembly and contributed to the old stellar populations of the inner-stellar halo. In general, the evidence for a high [Si/Fe] peak might plausibly be explained by a previous accretion event in the Milky Way. However, the level of Al enhancement observed in this new sub-population provide evidence that such stars are not from a dissolved dwarf galaxy \citep[see e.g.,][]{Hasselquist2019}.

 \subsection{A globular cluster origin} 
 
\begin{figure*}[t]
\begin{center}
\includegraphics[height = 13 cm]{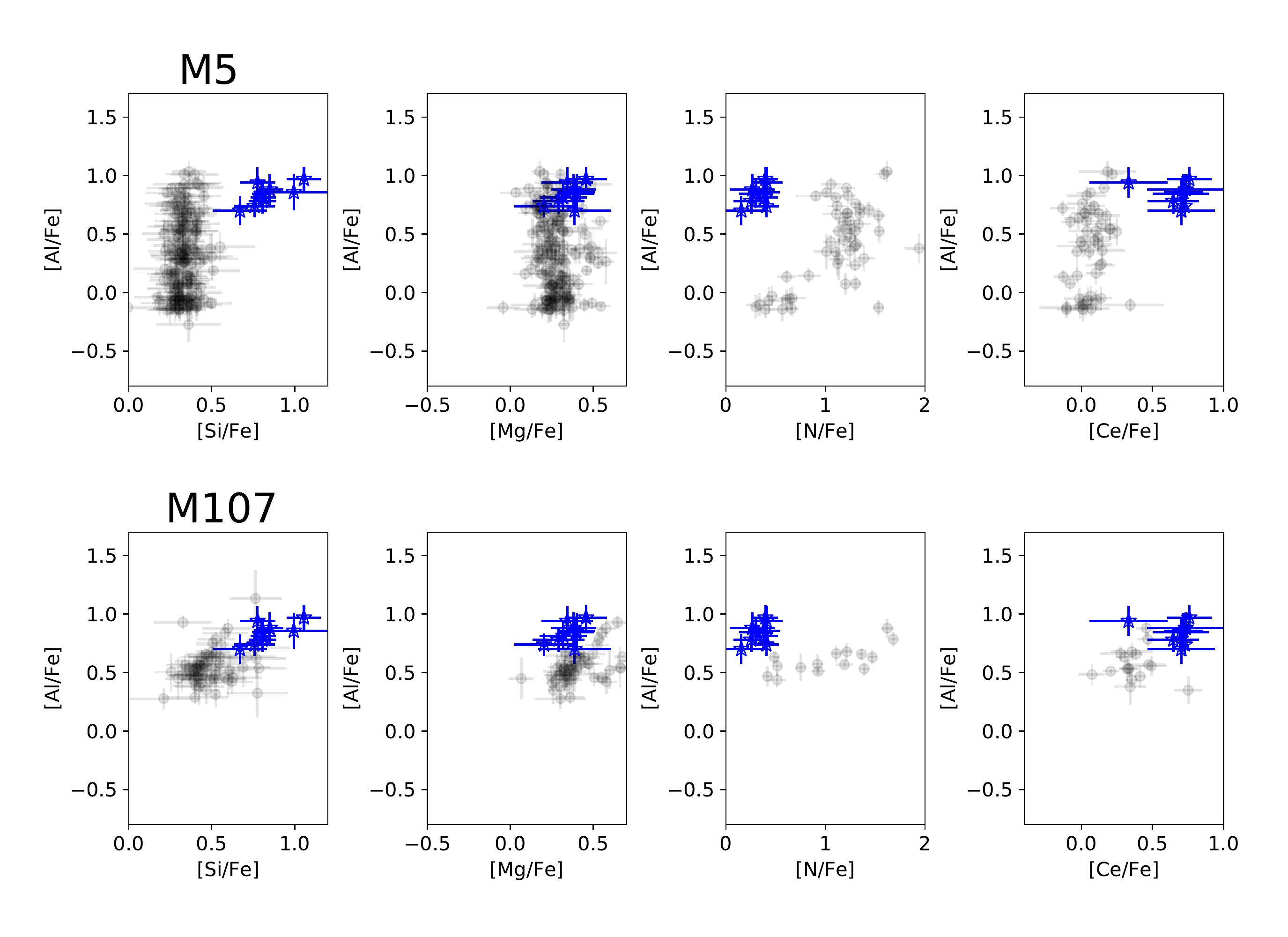}\\
\caption{
{M5 and M107 cluster stars from \citep{Masseron2019} are shown as filled grey dots in the Si-Al, Mg-Al, N-Al, and Ce-Al planes. The crosses indicate typical error bars. The Si-Al-rich stars are indicated with blue 'star' symbols.}
}
\label{fig3}
\end{center}
\end{figure*}

We speculate that some of the silicon-rich stars found in the inner Galaxy could be the remnant of a tidally disrupted in-falling GCs. Figure \ref{fig3} presents some the light and \textit{s}-process element patterns of the eleven star candidate pool identified of Si enhanced stars compared with those obtained from a run of \texttt{BACCHUS} for GCs stars \citep[see eg.,][]{Masseron2019} at the similar metallicity as our sample. Interestingly, similar Silicon enhancements have been detected in Galactic GCs \citep[][]{Masseron2019} Also in these cases, abundances of elements other than Silicon are similar to the observed for similar giant-stars. Remarkably, most of the newly identified silicon-rich field stars exhibit light-/heavy-element abundances similar to the typical Galactic GCs. It provides some clues either for the uniqueness of the progenitor stars to GCs, and is very likely that whatever process is responsible for such level of [Si/Fe] enhancement observed in the Galactic field appear to be similar from whatever caused the unusual stellar populations in GCs. Similar to our silicon-rich stars,  some Galactic GCs also exhibit interesting enrichment in \textit{s}-process elements \citep[e.g.,][]{Masseron2019} and anomalously high levels of Aluminum, Silicon, Cerium, and modest enrichment in Magnesium. In fact, most of the stars in our sample replicate (or exceed) the extreme abundance patterns of Galactic GCs, highlighting the uniqueness of these environments, as seen in Figure \ref{fig3}.  That is because it is very likely that silicon-rich stars are proposed to have been gestated in similar molecular clouds as Galactic GCs, thus sharing their environmental origin. However, it is less obvious in the [N/Fe]--[Al/Fe] plane, where the bulk of stars in Galactic GCs environments follow a clear correlation between [N/Fe] and [Al/Fe] (depending on the cluster or data set), this plot indicates that the GCs stars follow a slightly different N-Al enrichment that run parallel to those observed in the silicon-rich stars. It is important to note that silicon-rich stars were indeed discovered in GCs and in the inner stellar halo of the Milky Way (this study), but it is noteworthy that none so far has been identified in the disk of the Galaxy. 
 
 \subsection{Polluter candidates} 
 
 The origin of silicon-rich stars can be ascribed to different processes. If we allow a brief venture into the realm of speculation, it could be possible that mildly metal-poor giant-stars with Si overabundance could also originate through gas already strongly enriched from massive super novae (M$>20$ M$_{\odot}$), pair-instability supernova (PISNs), hypernovae (with stellar progenitors masses greater than $>140$ M$_{\odot}$), and/or faint supernovae \citep[see][]{Nomoto2013}, however many of these events involve the total destruction of the star, with not remnant left behind \citep[][]{Kemp2018}. Thus, the anomalously high levels of [Si/Fe] may be a result of the huge masses of processed material released from explosive nucleosynthesis of core-collapse supernovae (SNe II) events, which can explain the abundance feature of the silicon-rich giants. However, with the limited chemical species available in this study a single source of pollution is probably not enough to fit all the multi-elements observations. More stringent conclusions will be drawn when the full set of light-/heavy-elements will be available.
 
 Lastly, combining the absence of radial velocity variation ($RV_{SCATTER}<$ 0.5 km s$^{-1}$ as listed in Table \ref{table1}), and Silicon overabundance, does not provide enough information to support any hypothesis of such objects possibly being formed in a binary system. 
 
Conversely, the population of silicon-rich GC stars have been understood by invoking an internal production of Si, which is likely due to the increased leakage from Mg-Al cycle on $^{28}$Si, i.e., when the $^{27}$Al(p,$\gamma$)$^{28}$Si reaction takes over $^{27}$Al(p,$\alpha$)$^{24}$Mg a certain amount of $^{28}$Si, thus produced by proton-captures \citep{Karakas2003}. This overproduction of Si occurs typically at temperatures $>$ 100 MK \citep{Arnould1999}, in a similar way as have been recently detected in GCs stars \citep[][]{Masseron2019, Carretta2019}. The fact that the leakage from Mg-Al cycle on $^{28}$Si is apparently confined to GC environments implies that the process responsible for field stars with enhanced silicon may be tied to a certain epoch in the Milky Way's evolution similar to GCs stars at similar metallicity.
 
 \section{Concluding remarks}
 
 In conclusion, we find evidence of an unique collection of eleven mildly metal-poor giant-stars that display with an unexpected and exceptional anomalously high levels of  [Si/Fe] ($>+0.6$, which we henceforth refer to as silicon-rich stars), peaking at [Fe/H]$\sim-1.07$. In principle, this new sub-population form a distinct track in abundance plane, different from the Milky Way's field stars, supporting the identification of a separate origin. Here, we speculate that such stars could be former members GCs that still exist, or ancient clusters that were tidally disrupted. Whether these objects are mostly some of the oldest stars in the Milky Way is still yet to be determined. 
 
Furthermore, our dynamical study indicate that most of the silicon-rich stars selected so far possess halo-like and retrograde orbits passing through the bulge and disk of the Milky Way, which suggest that silicon-rich stars among field stars are either good candidates to be escaped GCs stars and/or the debris clues of GCs born in dwarf galaxies accreted to the Milky Way and that eventually got dissolved, spraying its stars across the Milky Way halo \citep[][]{Zinn1993, Majewski2012, Kruijssen2019}, however in this second . We speculate that this new stellar sub-population we discovered in the inner stellar halo may thus be the tracers of a global phenomenon where these stars were necessarily formed in Galactic GCs and were later lost to the field, such as $\omega$ Centauri \citep{Ibata2019}. However, a future inventory of the chemistry of these objects, in particular, the elements formed by neutron-capture processes, would hint at their origin, and possibly help confirm or refute the association with Galactic GCs stars.

\newpage

\acknowledgments

We acknowledge the anonymous referee for enlightening comments that greatly improved this Letter. We thank Szabolcs~M{\'e}sz{\'a}ros for helpful support computing the photometry $T_{\rm eff}$. J.G.F-T is supported by FONDECYT No. 3180210. T.C.B. acknowledge partial support for this work from grant PHY 14-30152; Physics Frontier Center / JINA Center for the Evolution of the Elements (JINA-CEE), awarded by the US National Science Foundation. B.T. acknowledges support from the one-hundred-talent project of Sun Yat-Sen University. SLM acknowledges funding from the Australian Research Council through Discovery grant DP180101791, and from the UNSW Scientia Fellowship program. Parts of this research were conducted by the Australian Research Council Centre of Excellence for All Sky Astrophysics in 3 Dimensions (ASTRO 3D), through project number CE170100013. SV gratefully acknowledges the support provided by Fondecyt reg. n. 1170518. A.P-V acknowledges a FAPESP for the postdoctoral fellowship grant no. 2017/15893-1 and the DGAPA-PAPIIT grant IG100319. J.G.F-T is grateful to Friedrich Anders for his precious help with \texttt{StarHorse}.

\texttt{BACCHUS} have been executed on computers from the Utinam Institute of the Universit\'e de Franche-Comt\'e, supported by the R\'egion de Franche-Comt\'e and Institut des Sciences de l'Univers (INSU). 

Funding for the \texttt{GravPot16} software has been provided by the Centre national d'\'etudes spatiales (CNES) through grant 0101973 and UTINAM Institute of the Universit\'e de Franche-Comt\'e, supported by the R\'egion de Franche-Comt\'e and Institut des Sciences de l'Univers (INSU). Simulations have been executed on computers from the Utinam Institute of the Universit\'e de Franche-Comt\'e, supported by the R\'egion de Franche-Comt\'e and Institut des Sciences de l'Univers (INSU), and on the supercomputer facilities of the M\'esocentre de calcul de Franche-Comt\'e. 

Funding for the Sloan Digital Sky Survey IV has been provided by the Alfred P. Sloan Foundation, the U.S. Department of Energy Office of Science, and the Participating Institutions. SDSS- IV acknowledges support and resources from the Center for High-Performance Computing at the University of Utah. The SDSS web site is www.sdss.org. SDSS-IV is managed by the Astrophysical Research Consortium for the Participating Institutions of the SDSS Collaboration including the Brazilian Participation Group, the Carnegie Institution for Science, Carnegie Mellon University, the Chilean Participation Group, the French Participation Group, Harvard-Smithsonian Center for Astrophysics, Instituto de Astrof\`{i}sica de Canarias, The Johns Hopkins University, Kavli Institute for the Physics and Mathematics of the Universe (IPMU) / University of Tokyo, Lawrence Berkeley National Laboratory, Leibniz Institut f\"{u}r Astrophysik Potsdam (AIP), Max-Planck-Institut f\"{u}r Astronomie (MPIA Heidelberg), Max-Planck-Institut f\"{u}r Astrophysik (MPA Garching), Max-Planck-Institut f\"{u}r Extraterrestrische Physik (MPE), National Astronomical Observatory of China, New Mexico State University, New York University, University of  Dame, Observat\'{o}rio Nacional / MCTI, The Ohio State University, Pennsylvania State University, Shanghai Astronomical Observatory, United Kingdom Participation Group, Universidad Nacional Aut\'{o}noma de M\'{e}xico, University of Arizona, University of Colorado Boulder, University of Oxford, University of Portsmouth, University of Utah, University of Virginia, University of Washington, University of Wisconsin, Vanderbilt University, and Yale University.

This work has made use of data from the European Space Agency (ESA) mission Gaia (\url{http://www.cosmos.esa.int/gaia}), processed by the Gaia Data Processing and Analysis Consortium (DPAC, \url{http://www.cosmos.esa.int/web/gaia/dpac/consortium}). Funding for the DPAC has been provided by national institutions, in particular the institutions participating in the Gaia Multilateral Agreement.


\end{document}